\documentclass[12pt]{article}
\newcommand{\be}{\begin{equation}}
\newcommand{\ee}{\end{equation}}
\newcommand{\ba}{\begin{eqnarray}}
\newcommand{\ea}{\end{eqnarray}}

\begin{document}

\title{Rigidity of cosmic acceleration in a class of k-essence cosmologies}
\author{ \large   Ruth Lazkoz \\
{\it \small Fisika Teorikoa, Zientzia eta Teknologiaren Fakultatea,}
\\
{ \it \small Euskal Herriko Unibertsitatea, 644 Posta Kutxatila, 48080 Bilbao, Spain}\\
}
\date{}
\maketitle
\begin{abstract}

We study the structural stability of a cosmic acceleration (inflation) in a class of k-essence cosmologies  against changes in the shape of the potential. Those models may be viewed
as generalized tachyon cosmologies and this analysis extends previous results
on the structural stability of cosmic acceleration in tachyon cosmologies. 
The study considers both phantom  and non-phantom cases.
The concepts of rigidity and fragility are defined through a condition on the functional form of the Hubble factor. Given the known result of the existence of inflationary (non-phantom) and super-inflationary (phantom) attractors we formulate the question of their structural stability.  We  find that  those attractors are rigid in the sense that they never change as long as the conditions for inflation or super-inflation are met.  
\end{abstract}

\maketitle
\titlepage
 \section{Introduction}
Even though WMAP data \cite{WMAP}
 support the inflationary paradigm , no completely satisfactory theoretical explanation of that phenomenon has been found yet.  Some works in this direction have attempted to find an answer in string theory, and as a result 
tachyonic inflation has been put forward \cite{sen}. The idea strongly relies in the possibility of describing tachyon condensates in string theories
in terms of perfect fluids. Such models are derived from the
factorizable  Born-Infeld  Lagrangian
${\cal L}=-V(\phi)\sqrt{1-\dot\phi^2}$
which was associated with the tachyon  by computations in boundary string
field theory \cite{BSFT}. Such Lagrangian also arises in open bosonic string theory \cite{fratse} and is a key ingredient in the effective
theory of D-branes \cite{Leigh}. The energy density and pressure of cosmologies derived from that Lagrangian  will not be canonical in the sense that they will not depend linearly on the  kinetic energy.

 In general, cosmological models derived from  Lagrangians with non-cano\-nical kinetic terms are dubbed k-essence cosmologies. Such models have not only been considered for the modest goal of describing inflation, but also for the most ambitious purposes of describing late  time acceleration induced by dark energy or even for the unification of dark energy and dark matter. Nevertheless, here we will concentrate on issues to do with inflation only.

Construction of cosmological models  involves a great deal of idealization, and k-essence models are no exception. It may turn out that conclusions or results  depend strongly on the chosen values and number of free parameters in the potential. 
There are two main reasons why cosmological modeling is never perfect: First,
simplifying assumptions made in common practice are chosen for technical convenience and have little to do with observations; second,  cosmological observations  depend on the model  and so add to the possible errors, and  thus, even models based on observation will necessarily  be imperfect. To make things worse, the existence of a large number of noticeably different mathematical models which agree so well with the data make a very bad case of the empirical methods used nowadays in physics \cite{golda}.  

As put forward many years ago by Andronov and Pontryagin \cite{stst}, in one wishes to build satisfactory models of real phenomena (not just in cosmology) then one should try and make ensure their structural stability, that is,  one should provide predictions qualitative independent of perturbations. In addition, it has been claimed \cite{coley} that the concepts of rigidity and fragility seem to be important for most cosmological models, and that structural fragility might be the suitable theoretical setup for cosmology \cite{tavakol}.

Turning back to  inflation driven by k-essence, we address the problem of its  rigidity or fragility  from the perspective of a class of models which can be viewed as generalizations of the conventional tachyon cosmologies. The relation between these generalized tachyon cosmologies and the conventional ones is that there is a relation of proportionality between the effective speed of sound of those two classes of k-essence.

In analysis of structural stability of inflation in such models we follow closely an approach devised for standard scalar fields in \cite{Lidsey}, as we did in our previous
work regarding conventional tachyon cosmologies  \cite{agulaz} (this technique has also been recently applied to study the structural stability of inflation with phantom fields). The method allows identifying regions where
the attractor solutions change, thus indicating fragility in the system.  

The evolution of our models, and of any other FRW cosmology with a single matter component, is given by a two-dimensional set of equations. This is particularly interesting, because according to the Peixoto theorem for as close as wished to any two-dimensional
dynamical system there exists another one which is structurally stable. Therefore, it is justified to require the structural stability of inflation of FRW cosmologies with a single matter component. Note that at the end of the day inflation will only represent a realistic
phenomenon if it shared by both a particular model and by slightly perturbed versions of it. 

The plot of the paper is as follows. In Section 1 we outline the main equations, in Section 2, we prove the existence of inflationary attractors, in Section 3 we prove the rigidity of inflation in the models under discussion, and finally in Section 4 we draw our main conclusions.

\section{Basic setup}
The Einstein equations for a flat ($k=0$) Friedmann-Robertson-Walker (FRW) cosmological
model with a perfect fluid with energy density $\rho$ and pressure $p$
\ba
&&3H^2=\rho\\
&&2\dot H=-(\rho+p)\label{hdot},
\ea
and they lead to the conservation equation
\be
\dot\rho+3H(\rho+p)\label{conserv},
\ee
where overdots denote differentiation with respect to cosmic time $t$, $H\equiv  \dot a/a $ is the 
Hubble parameter, and $a$ is the synchronous scale factor. 

Let us recall now that conventional tachyon cosmologies \cite{tachyon} are a class of k-essence cosmologies \cite{k-dark}. If the tachyon field is $\phi$ and the self-interaction potential is 
$V(\phi)$ we have 
\ba
&&\rho=\frac{V(\phi)}{\sqrt{1\mp\dot\phi^{2}}},\\
&&p=-V(\phi)\sqrt{1\mp\dot\phi^{2}}.
\ea
where here and throughout the upper sign will correspond to non-phantom models and
the lower one to phantom ones. Recall that for phantom models $\rho>0$ and $\rho+p<0$. Note as well that switching from  the non-phantom to the phantom case only takes a Wick rotation of the field \cite{aguchilaz}.

The stability of the k-essence with respect to small wavelength
perturbations requires that the square of the effective sound speed given by
the definition \cite{kinfper}
\be
c_s^2=\frac{\partial p/\partial \dot\phi^2}{\partial\rho/\partial \dot\phi^2}
\ee
be positive. However, in  \cite{car} it was shown that a positive
sound speed is not a sufficient condition for the theory to
be stable. In the case of the conventional tachyon cosmologies one has
\be
c_s^2=1\mp\dot\phi^2,
\ee
and combining this result with the aforementioned requirement on $c_s^2$, in \cite{Chimento} the author
looked for the families of k-essence cosmologies  such that their sound speed is proportional to that
 of the conventional tachyon. In that reference it was  found that if $r$ is an arbitrary parameter, then k-essence models with a sound speed $2r-1$ times smaller
 than  conventional tachyon models correspond to having
\ba
\rho=V(\phi)(1\mp\dot\phi^{2r})^{\frac{1-2r}{2r}},\\
p=-V(\phi)(1\mp\dot\phi^{2r})^{ \frac{1}{2r}}.
\ea
Accordingly,  Eqs. (\ref{hdot})  and (\ref{conserv}) become respectively
\be
\frac{2\dot H}{3H^2}=\mp\dot\phi^{2r}\label{evol}
\ee
and
\be
\frac{V,_{\phi}}{V}\pm 3H\dot\phi^{2r-1}+(2r-1)\frac{\dot\phi^{2(r-1)}{\ddot \phi}}{1\mp\dot\phi^{2r}}=0 \label{k-g}
\ee

Now, our purpose is to look for correspondences between the space of scale factors and that of inflationary potentials in the setup of this class of k-essence cosmologies. We will set the discussion in a general framework (valid for any potential)
which is an adaptation of Lidsey's
approach \cite{Lidsey}.

Nevertheless, before we go on we have to study  whether inflationary solutions are asymptotically stable, i.e. whether inflationary attractors exist. 
Typically, for inflation to proceed it is necessary that the energy density be dominated by the potential energy, which is equivalent to having a negligible kinetic energy. In such situation, it will be possible to discard the term proportional to $\ddot\phi$ in  (\ref{k-g}). As a consequence inflationary attractors must be solutions satisfying
$\gamma=-2\dot H/3H^2\approx{\rm cons}$.

The stability of k-essence models with constant barotropic index was studied in \cite{ChiLaz}, where it was shown that stability required $\gamma<1$, so  accelerated 
and super-accelerated (phantom) solutions are attractors. This proves the existence of inflationary attractors.

Having proved, the existence of generalized tachyon cosmologies with accelerated expansion, we can now move on to see how the features of the attractors depend on the form of the potential.

\section{Structural stability}
We start off by rewriting the equations of motion, using the Hamiltonian formalism. The Friedmann constraint can be cast in the following form:
\be
\left(3H^2\right)^{2r}=V^{2r}\left(1\mp\left(\mp\frac{2H'}{3 H^2}\right)^{\frac{2r}{2r-1}}\right)^{1-2r}\label{hamilton},
\ee
where here and throughout primes denote differentiation with respect to $\phi$. 

Clearly, the task of obtaining most of the important results to do with  accelerated expansion gets
simplified upon using the Hamiltonian formalism. Specifically, the power and appeal of this technique lies in the fact that it allows to
consider $H(\phi)$, rather than $V(\phi)$ as the fundamental quantity to be specified.

Solutions to (\ref{hamilton}) can be labelled by means of a parameter $p$, so that we have $H(\phi(t),p)$. The value of $p$ is fixed unambiguously once the initial conditions have been
chosen. The corresponding expression for the scale factor will be
\be
a(\phi(t),p)=a_i \,{\rm exp}\left(\int_{\phi_i}^{\phi}d\tilde \phi H(\tilde \phi,p)\left(\mp\frac{\partial H(\tilde\phi,p)/\partial \tilde \phi}{3H^2(\tilde\phi,p)/2}\right)^{-1/(2r-1)}\right),
\ee
where $a_i$ and $\phi_i$ are constants of integration. 

Let us consider now two solutions $H(\phi,p+\Delta p)$ and $H(\phi,p)$, under the requirement that
they are very close together in the corresponding space, i.e. $\vert\Delta p\ll 1\vert$.  
We then have 
\be
H(\phi,p+\Delta p)-H(\phi,p)\approx\left(\partial H/\partial p\right)_\phi\Delta p\,.\ee
By differentiating  (\ref{hamilton}) with respect to $p$, and combining the result 
with the evolution equation (\ref{evol}) and the definition of $H$ we get
\be
H(\phi,p+\Delta p)-H(\phi,p)\propto a^{-3}(\phi,p)\Delta p\label{eq:ha}\,,
\ee
which coincides exactly with the formula obtained in \cite{agulaz} for conventional tachyon cosmologies. 
 
As a consequence of  (\ref{eq:ha}), the differences between not very different solutions disappear as the models evolve, which means they all approach some 
attractor solution $H(\phi)$. However, the attractor may not be the same for all values
of the potential parameters; put another way, the system may be fragile around the point
at which the attractors change. In order to check whether that is the case, one defines
the quantity
\be
F\equiv\left\vert\frac{H(\phi,p+\Delta p)}{H(\phi,p)}-1\right\vert\,.
\ee
Equation (\ref{eq:ha}) shows that in an expanding universe $F\to0$ as time grows, but the form of the attractor
may vary if $\partial F/\partial \phi$ changes sign. Note that $F$ can go to zero for $\partial F/\partial \phi>0$ or for $\partial F/\partial \phi<0$, but not for both. Thus, 
if  $\partial F/\partial \phi=0$ for some value of the parameters the system will be said
to be fragile around that very value. For convenience, we  look for sign changes
in $\partial \log F/\partial \phi$ instead of $\partial  F/\partial \phi$. It can be seen
that
 \be
 \frac{\partial \log F}{\partial \phi}=-3 H(\phi,p)\left(\mp\frac{3H^2(\phi,p)}{2H'(\phi,p)}\right)^{\frac{1}{2r-1}}-\frac{H'(\phi,p)}{H^2(\phi,p)}.
 \ee
Therefore, the fragility condition is
\be
\pm\left(\mp\frac{2H'}{3H^2}\right)^{\frac{2r}{2r-1}}=2\label{fragility},
\ee
which, as will be shown immediately,  is all we need to answer the question
of whether cosmic acceleration is rigid in the models under consideration. 
In flat FRW models the condition for accelerated expansion  is $- \dot H/H^2<1$ and in the case of generalized tachyon cosmologies it can be seen to be equivalent to
\be
\pm\left(\mp\frac{2H'}{3H^2}\right)^{\frac{2r}{2r-1}}<1.
\end{equation}
Clearly,  rigidity of cosmic acceleration or super-acceleration  
follows automatically from the latter. 

\section{Conclusions}
Realistic models of the universe are believed to have to be structurally stable, in the sense that the if we have  a particular model with desirable properties then its  slightly perturbed versions must also display the same property.  Models of a flat homogeneous universe with a single matter component are described by a two-dimensional
set of evolution equations, and the requirement of the structural stability of
two-dimensional dynamical systems is justified by the Peixoto theorem. 

In this paper we have proved  the rigidity (i.e., the structural stability) of cosmic acceleration in a class of  k-essence cosmologies by using a procedure that allows spotting changes in the attractor solutions just by checking at the value of some function of a simple function of the Hubble factor and its derivative with respect to the tachyon field. Since those  k-essence cosmologies can be viewed as generalized tachyon cosmologies the rigidity of conventional tachyonic inflation (which was discussed in a previous work \cite{agulaz}) is just a particular application of the discussion in this paper.
All along we have considered non-phantom and phantom cases, for the former we have proved
the rigidity of inflation, whereas for the latter  we have proved
the rigidity of super-inflation.

\section*{Acknowledgments}
Thanks to J.M. Aguirregabiria and L.P. Chimento for conversations.  Support from the University of the Basque Country through research grant UPV00172.310-14456/2002,   from former the Spanish Ministry of Science jointly with FEDER funds through research grant  BFM2001-0988 and
from the Spanish Ministry of Science and Education through research grant FIS2004-01626 is also acknowledged. 
 
\end{document}